# Assessing Phenotype Definitions for Algorithmic Fairness

Tony Y. Sun, MA[1], Shreyas A. Bhave, MA[1], Jaan Altosaar, PhD[1], Noémie Elhadad, PhD[1]
[1]Columbia University, New York, New York

**Abstract**

*Phenotyping is a core, routine activity in observational health research. Cohorts impact downstream analyses, such as how a condition is characterized, how patient risk is defined, and what treatments are studied. It is thus critical to ensure that cohorts are representative of all patients, independently of their demographics or social determinants of health. In this paper, we propose a set of best practices to assess the fairness of phenotype definitions. We leverage established fairness metrics commonly used in predictive models and relate them to commonly used epidemiological metrics. We describe an empirical study for Crohn's disease and diabetes type 2, each with multiple phenotype definitions taken from the literature across gender and race. We show that the different phenotype definitions exhibit widely varying and disparate performance according to the different fairness metrics and subgroups. We hope that the proposed best practices can help in constructing fair and inclusive phenotype definitions.*

**Introduction**

When conducting an observational health study, one of the core, routine tasks researchers must address is defining the study population. If the population of interest is a set of patients with a particular manifestation of disease, this task is referred to as phenotyping. Phenotype definitions select patients into disease cohorts that are used to improve our collective knowledge about a particular condition, including fundamental epidemiological queries (e.g., quantifying incidence of disease overtime)[1], a risk estimation and prediction questions (e.g., identifying risk factors for stroke)[2] and comparative effectiveness studies (e.g., comparing diuretics vs ace-inhibitors for treating hypertension)[3]. High-impact research using phenotypes eventually impacts policy-making about potential medical treatments, and consequently the health of populations[3–5]. Thus, it is critical to evaluate whether phenotype definitions adequately represent all patients in a population.

Despite our best efforts as a research community to reduce bias in phenotype construction, phenotype definitions are still subject to multiple potential sources of bias; we illustrate here three such biases. **Diagnosis bias** prevents (or delays) disease diagnosis, often because of differences in initial presentation of disease across sub-groups. For instance, a meta-analysis of acute myocardial infarction symptom literature finds that men presenting with acute myocardial infarction are more likely to complain of chest pain, while women are more likely to complain of other forms of pain. A phenotype definition designed with chest pain as the primary presenting symptom may thus under-represent women[6]. **Treatment bias** prevents (or delays) appropriate medical treatment for individuals in certain groups. For instance, in a study of patients at the VA, evidence shows that Black patients are less likely to be prescribed cardio-protective drugs (beta blockers, statins, and ACE inhibitors) than white patients[7]. A phenotype definition for hyperlipidemia for instance might require proof of treatment. Under this condition, the phenotype might under-represent Black patients. Lastly, **access to care biases** are systemic issues that prevent patients from getting into the healthcare system. As such, a phenotype definition that requires the disease diagnosis to occur in a particular out-patient setting could under-represent patients that primarily rely on emergency visits for their regular care.

As standards in observational health data have emerged, the research community has acknowledged the need to standardize and validate phenotype definitions across multiple research sites (like in, for instance, the eMERGE network[8]. This practice has helped avoid potential biases in patient selection with respect to institution-specific documentation practices and geographical location idiosyncrasies. However, to date, there is no standard approach to assess the fairness of a phenotype definition beyond institutional care documentation differences.

In this paper, we develop and propose a set of **best practices for evaluating phenotype definitions** by bridging the evaluation measures used in epidemiological literature with those used in the algorithmic fairness and machine learning literature. We hope these practices can help make the adoption and reporting of fairness metrics in observational health studies standard practice. As **illustrative scenarios**, we assess the fairness of several highly-cited phenotypes for Crohn's disease and diabetes type 2 with respect to gender (women, men) and race (Black, white) and show that phenotype definitions have varying performance characteristics, revealing real-world tradeoffs.

**Related Work**

The related work fall broadly into two categories: (1) evaluation and validation of phenotypes and (2) algorithmic fairness in healthcare.

*Evaluation and Valuation of Phenotype Definitions*

Rule-based and unsupervised phenotyping algorithms are primarily evaluated via clinical adjudication of patient records. The most common approach is to take a random sample of subjects identified by the phenotype and recruit one or more clinical experts to assess whether each subject meets the criteria for inclusion. Such an analysis typically yields a single number, an estimate of the positive predictive value (or precision) of the phenotype, which is calculated by comparing the gold standard clinical labels against the predicted labels from the phenotype algorithm. Related performance metrics such as sensitivity and specificity are rarely assessed, as they require clinical review of a much larger set of subjects (i.e., subjects identified as having the disease and not).

Previous literature on the subject of phenotype evaluation has already identified that phenotype validation, if it occurs at all, is rare and often incomplete[9]. This evaluation gap is persistent across numerous disease phenotypes. For example, in a systematic review of myocardial infarction phenotypes, researchers reported that of 33 validation studies, only 11 reported sensitivity and 5 reported specificity (all provided estimates for positive predicted value)[10]. This trend of providing precision but rarely estimating sensitivity and specificity is borne out in review of disease phenotypes for atrial fibrillation and stroke[11,12].

The same trends are also present in the two diseases (Type 2 Diabetes and Crohn's) we focus on in this paper. For Crohn's disease, most publications which leverage rule-based phenotypes do not mention any form of clinical validation[13,14]. Among the phenotypes we leveraged in this paper, three report sensitivity and specificity along with positive predictive value[15–17], while the others only report precision[18]. For Type 2 Diabetes, we also find that many papers do not mention any clinical validation[19,20]. Among the phenotypes we use, two of them report sensitivity, specificity, and precision[21,22].

Since phenotypes in most cases are not evaluated even using population-wide validation metrics such as sensitivity and specificity, it is evident that the vast majority of published phenotypes do not report these statistics within sub-groups or stratified by protected classes. Swerdel et al. proposed a general approach to evaluating a phenotype definition and estimate these population-level metrics, even in the absence of gold standard, but does not provide guidance for assessing its fairness[9].

*Algorithmic Fairness in Healthcare*

Recently, algorithms used to make healthcare decisions have come under greater scrutiny for potentially being biased against certain protected classes such as race and gender. For example, Obermeyer et al. demonstrated that a commercial algorithm used to identify and select at-risk patients was biased against Black patients because the algorithm was primarily focused on cost minimization[23]. McCradden et al. emphasizes that existing performance metrics such as sensitivity and specificity might "camouflage" persistent health inequities, and suggests reporting group-specific performance metrics for algorithms trained on fundamentally biased healthcare data[24].

Ethical machine learning in healthcare requires considering how biases might be introduced at all levels of the experimental design process including problem selection, data collection, outcome definition, and algorithm development[25]. While much research has focused on bias in downstream tasks such as clinical outcome prediction, less emphasis has been placed on upstream steps such as data collection and outcome definition. Phenotyping algorithms are one such early step in many observational studies for which there is no standard for assessing whether these biases exist. Given the broad use of phenotypes in observational health research, we argue that biases introduced at the level of phenotypes harbor the risk of exacerbating existing health disparities by influencing clinical guidelines and public policy.

**Best Practices for Assessing the Algorithmic Fairness of Phenotypes**

We propose a set of best practices for the research community to use when constructing and assessing phenotype definitions. These best practices are centered around the use of fairness metrics commonly cited in the fairness in machine learning literature. We bridge these fairness metrics with commonly used epidemiological measures, enabling researchers to interpret tradeoffs more easily. First, we detail the interpretation of fairness metrics as common epidemiological measures (Figure 1). Subsequently, we enumerate best practices for how to use these metrics when developing a phenotype. We hope that using these best practices will enable researchers to be transparent, intentional and explicit about the assessment and construction of their phenotypes.

**Fairness Metrics: An Epidemiological Perspective**

To assess the impact of phenotype algorithms on subgroup inclusion, we rely on fairness metrics that are commonly applied to supervised models. As notation, let $A \in \{0, 1\}$ refer to a protected attribute of a patient which could be a binary variable, like the gender (woman, man) or race (Black, white) of a patient. Let $\hat{Y} \in \{0,1\}$ be the predicted output from a phenotyping function, for exclusion or inclusion in a disease cohort. Further, let $P_0(\hat{y}) = P(\hat{y}|A = 0)$ be the predicted output of a phenotyping function given a protected attribute. Below, we define and compare how these metrics can map to pre-existing epidemiological measures.

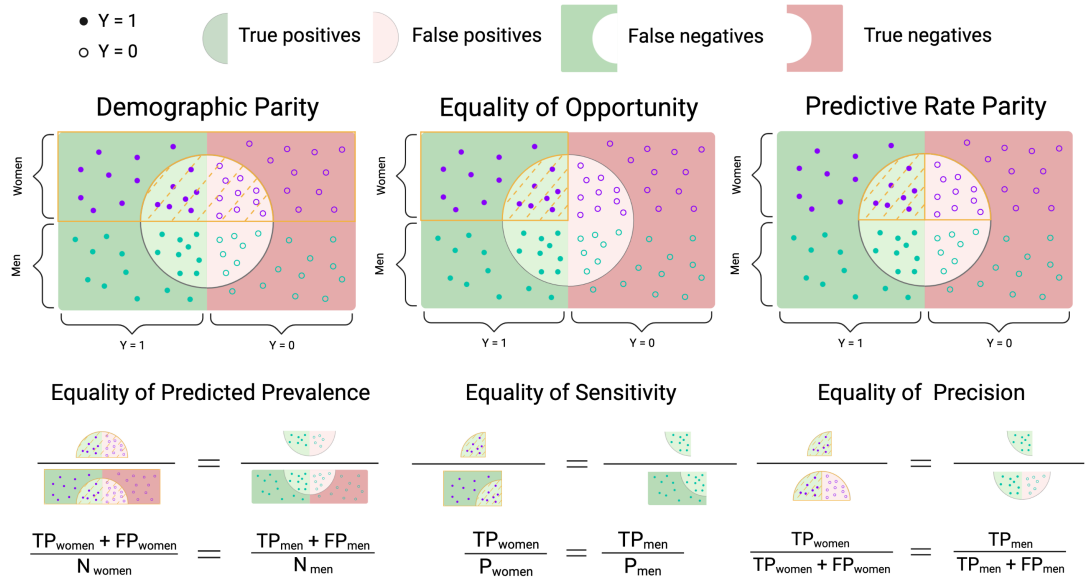

**Figure 1. Algorithmic fairness metrics mapped to existing epidemiological measures stratified by a protected class such as gender.** We visualize three fairness metrics, by treating phenotyping algorithms as classifiers. In taking this perspective, we can equate demographic parity to equality of predicted prevalence; equality of opportunity to equality of sensitivity; and predictive rate parity to equality of precision.

*Demographic Parity as Equality of Predicted Prevalence*

Also called "independence" or "statistical parity" in the fairness literature, demographic parity is the difference in the proportion for each protected class that receives the positive (and negative) outcome. In epidemiology, minimizing demographic parity would be equivalent to asserting that, among patients diagnosed with the disease, protected classes should have the same prevalence among the diagnosed. Mathematically this translates to:

$$P_0(\hat{Y} = \hat{y}) = P_1(\hat{Y} = \hat{y}) \ \forall \ \hat{y} \in \{0,1\}$$

The basis of demographic parity is the federal four-fifths rule, which states that "a selection rate for any race, sex, or ethnic group which is less than four-fifths of the rate of the group with the highest rate will generally be regarded by the Federal enforcement agencies as evidence of adverse impact". For many diseases, we might *a priori* expect there to be differences in prevalence across protected categories, for reasons unrelated to disparate treatment - for example, scientific literature has identified biological reasons for why many autoimmune disorders are more prevalent among women than men[26]. Despite this, demographic parity can be an appropriate in some healthcare settings or for some phenotypes where strict non-discriminatory practices are desirable[27], such as phenotypes for inpatients visits or outpatient referrals for common conditions.

*Equality of Opportunity as Equality of Sensitivity*

Also called "separation" or "positive rate parity" in the fairness literature, equality of opportunity is achieved when the true positive rates of a model are equal across demographics. In epidemiology, maintaining equality of opportunity is equivalent to ensuring that the sensitivities within each protected class are equal. This translates to:

$$P_0(\hat{Y} = 1 | Y = 1) = P_1(\hat{Y} = 1 | Y = 1)$$

Equality of opportunity does not adjudicate false positives (or high precision) - it primarily captures whether the same proportion of patients across classes truly "had" the disease. To calculate the equality of opportunity, note that a "true" label is required. We approximate this ground-truth label using a silver standard by computing the majority vote label across multiple published phenotypes (see *Approach: Assessing Fairness of Crohn's Disease and Diabetes Phenotype Definitions*).

*Predictive Rate Parity as Equality of Precision*

Also called "sufficiency" in the fairness literature, predictive rate parity is achieved when the probability of the true labels given the predicted label is equivalent across classes. In epidemiology, maintaining predictive rate parity is equivalent to ensuring that phenotypes have equivalent precision within each protected class. This translates to:

$$P_0(Y = 1 | \hat{Y} = 1) = P_1(Y = 1 | \hat{Y} = 1)$$

and

$$P_0(Y = 0 | \hat{Y} = 0) = P_1(Y = 0 | \hat{Y} = 0)$$

Note that again that to calculate this a "true" label is required, which we discuss below.

**Using Fairness Metrics to Assess and Construct Phenotypes**

We now highlight steps researchers can take when building and assessing phenotypes using these fairness metrics:

1. **Enumerate protected classes.** The first step in the workflow is to simply enumerate all protected classes which are of interest in phenotyping a particular disease. Examples could include race (Black, white, etc.), gender (women, men, etc.), ethnicity (Hispanic, non-Hispanic, etc.), or age group (patients under 18, over 65, etc.).

2. **Identify priorities to optimize.** Next, prioritize optimizing a particular fairness metric before constructing a phenotype definition. For example, in certain scenarios an all-encompassing, highly sensitive definition for *all* protected classes may be desirable. On the other hand, a specific definition which exhibits high precision may be desirable in other cases.

3. **Construct phenotype.** The phenotype can now be designed with priorities in mind. The priorities will affect how broad inclusion criteria are and whether to incorporate diverse data types (specific medications, treatment regimes, etc.).

4. **Acquire gold/silver standard.** A gold or silver standard is required to assess fairness metrics. If a gold standard cannot be obtained on a subset of patients, a silver standard may be constructed using methods like PheValuator or majority vote across many commonly used phenotypes for the same disease[9].

5. **Compute fairness metrics.** Use the gold/silver standard to compute fairness metrics and assess the tradeoffs and their epidemiological interpretation. For example, perhaps the phenotype is poorly sensitive for Black patients under the equality of opportunity criterion compared to white patients.

6. **Revise phenotype definitions.** Based on the fairness metrics, revise the phenotype definition to attempt to mitigate any fairness gaps. For example, requiring continuous observation for a year or at least one inpatient stay may decrease the sensitivity of the phenotype for Black patients who have fewer interactions with the healthcare system. This inclusion criteria may then be dropped and then fairness metrics may be recomputed.

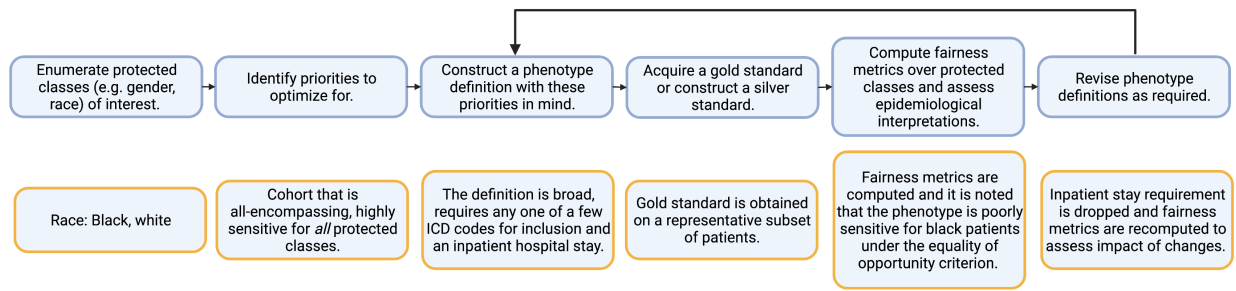

**Figure 2. Best Practices for Assessment and Construction of Fair Phenotypes.** We enumerate a sequence of steps that researchers can take to develop phenotypes with their specific concerns regarding biases across protected groups in mind. The steps are highlighted in the top blue boxes and an example of this workflow applied in practice is shown in the orange boxes.

In (Figure 2) above we provide the full workflow and an example scenario. We hope that the use of this workflow may aid researchers in (1) making their priorities for their phenotype explicit (2) being intentional about which protected groups are being considered in phenotype design (3) assessing the fairness tradeoffs intuitively in epidemiological language (4) iteratively redesigning their phenotypes to optimize for their priorities. We recommend that researchers document this iterative process and disseminate all relevant performance and fairness metrics associated with their phenotype.

**Approach: Assessing Fairness of Crohn's Disease and Diabetes Phenotype Definitions**

We illustrate the use of the proposed best practices on the following scenarios for two conditions: Crohn's disease and type 2 diabetes. We selected these two conditions, as they are well-studied diseases with publicly available phenotype definitions. For each condition, we compare multiple phenotype definitions and assess their fairness across two types of patient subgroups (gender and race). Each definition is derived from highly-used or highly-cited publications.

To assess the equality of opportunity and predictive rate parity fairness metrics, ground truth labels are needed about who has a disease. We approximate this using a silver standard, which we define as the group of patients that belong to a majority of phenotypes for a given disease. For example, in our Crohn's analysis using five distinct phenotype definitions, patients that belong to at least three of the five phenotype definitions are labeled as "true positives."

*Data source and phenotype source*

The different phenotypes are implemented using electronic health record data from NewYork-Presbyterian Hospital, a tertiary academic medical center that serves a heterogeneous patient population in New York City. The data are translated and standardized to the Observational Medical Outcomes Partnership (OMOP) common data model[28], which allows research to easily extend to other institutions using the same common data format. Phenotype definitions are taken from the Observational Health Data Sciences and Informatics (OHDSI) PhenotypeLibrary, which provides publicly-available phenotype definitions for various disease diagnoses, along with citations explaining their sourcing.

*Crohn's phenotypes*

We choose five Crohn's disease phenotypes from the OHDSI PhenotypeLibrary: (1) a commonly used OHDSI literature definition most recently included in the ongoing OHDSI Health Equity Research Assessment (HERA) study that was slightly modified based on existing literature[15], (2) the original heavily-cited Crohn's phenotype[15], (3-4) phenotype definitions from publications that have impacted clinical diagnosis or medical guidelines[17,18], and (5) another heavily-cited Crohn's phenotype[16]. Phenotype demographics are provided in (Table 1), while phenotype descriptions as well as the number of included concepts are provided in (Table 3).

**Table 1.** Crohn's disease phenotype demographics: The gender, race, and age demographic information for each Crohn's phenotype implemented.

|        |       | HERA (2021) | Thirumurthi (2010) | Ananthakrisnan (2013) | Stepaniuk (2015) | Benchimol (2014) |
|--------|-------|-------------|--------------------|-----------------------|------------------|------------------|
| Gender | Women | 4085        | 1670               | 6809                  | 532              | 1129             |
|        | Men   | 3222        | 1587               | 6124                  | 543              | 1025             |
| Race   | Black | 510         | 292                | 779                   | 101              | 173              |
|        | White | 3387        | 1174               | 5793                  | 483              | 1143             |
| Age    | ≤ 18  | 3.85%       | 0.03%              | 3.97%                 | 0.00%            | 6.69%            |
|        | 18-30 | 12.55%      | 6.42%              | 14.46%                | 9.86%            | 21.68%           |
|        | 30-60 | 37.26%      | 38.78%             | 42.27%                | 53.86%           | 40.20%           |
|        | 60-80 | 28.30%      | 27.72%             | 24.84%                | 18.33%           | 23.17%           |
|        | ≥ 80  | 19.02%      | 27.97%             | 15.20%                | 18.51%           | 8.82%            |

*Type 2 Diabetes phenotypes*

We choose three diabetes disease phenotypes from the publicly-available phenotypes listed on the OHDSI PhenotypeLibrary (1) a highly cited literature definition that requires specific diagnoses, medication, and hemoglobin A1c measurements[21], (2) a widely used OHDSI definition that proposed new therapeutic guidelines for hypertension prescriptions[3], and (3) a validated diabetes phenotype from the pheKB initiative designed to be portable across sites[22]. Phenotype demographics are provided in (Table 2) below, while phenotype descriptions as well as the number of included concepts are provided in (Table 4).

**Table 2.** Crohn's disease phenotype demographics: The gender, race, and age demographic information for each Crohn's phenotype implemented.

|        |       | Miller (2004) | LEGEND (2014) | PheKB (2012) |
|--------|-------|---------------|---------------|--------------|
| Gender | Women | 70,989        | 71,080        | 2,915        |
|        | Men   | 55,343        | 61,278        | 3,240        |
| Race   | Black | 15,515        | 16,676        | 729          |
|        | White | 42,150        | 39,588        | 2,010        |
| Age    | ≤ 18  | 0.54%         | 0.69%         | 0.50%        |
|        | 18-30 | 2.51%         | 1.87%         | 1.90%        |
|        | 30-60 | 25.83%        | 21.13%        | 27.58%       |
|        | 60-80 | 43.71%        | 43.53%        | 47.69%       |
|        | ≥ 80  | 29.40%        | 34.82%        | 24.32%       |

**Tables 3 and 4.** Crohn's disease phenotype demographics: The gender, race, and age demographic information for each Crohn's phenotype implemented.

Table 3: Crohn's Phenotype Definitions

| Phenotype | Definition |
|-----------|------------|
| HERA | ICD-9: 555.x; ICD-10: K50.x<br>no prior history of Crohn's<br>(59 unique codes) |
| Thirumurthi (2012) | ICD-9: 555.0, 555.1, 555.2, 555.9<br>Visits: in-patient/ER<br>Age: 18+<br>(5 unique codes) |
| Ananthakrisnan (2013) | ICD-9: 555.0, 555.1, 555.2, 555.9<br>(4 unique codes) |
| Stepaniuk (2015) | ICD-10: K50 w/o descendants<br>Visit: in-patient or ED<br>(31 unique codes) |
| Benchimol (2014) | ICD-9: 555.x; ICD-10: K50.x<br>no prior history of Crohn's<br>5 Crohn's diagnoses in 4 years<br>(59 unique codes) |

Table 4: Diabetes Type 2 Phenotype Definitions

| Phenotype | Definition |
|-----------|------------|
| Miller (2004) | Miller custom concept set<br>no history of diabetes<br>glycemic med., HbA1c measurement<br>(18,649 unique codes) |
| LEGEND (2019) | OHDSI custom concept set<br>diabetes med. (besides insulin)<br>at least 1 HbA1c measurement<br>(18,711 unique codes) |
| PheKB (2012) | PheKB custom concept sets<br>diabetes med. (besides insulin)<br>HbA1c measurement or glucose lab results<br>(19,017 unique codes) |

## Results

After empirically analyzing Crohn's and type 2 diabetes mellitus phenotyping definitions across the three fairness metrics, we identify examples of significant real-world trade-offs arising from phenotype construction. In (Figure 3) and (Figure 4) below we visualize the demographic parities, equality of opportunities, and predictive rate parities across gender and race for Crohn's patients and type 2 diabetes mellitus patients, respectively. Statistically significant differences between subgroups are assessed by calculating the difference in proportions using a two-sided proportion z-test, at significance level α = 0.05.

*Crohn's phenotypes*

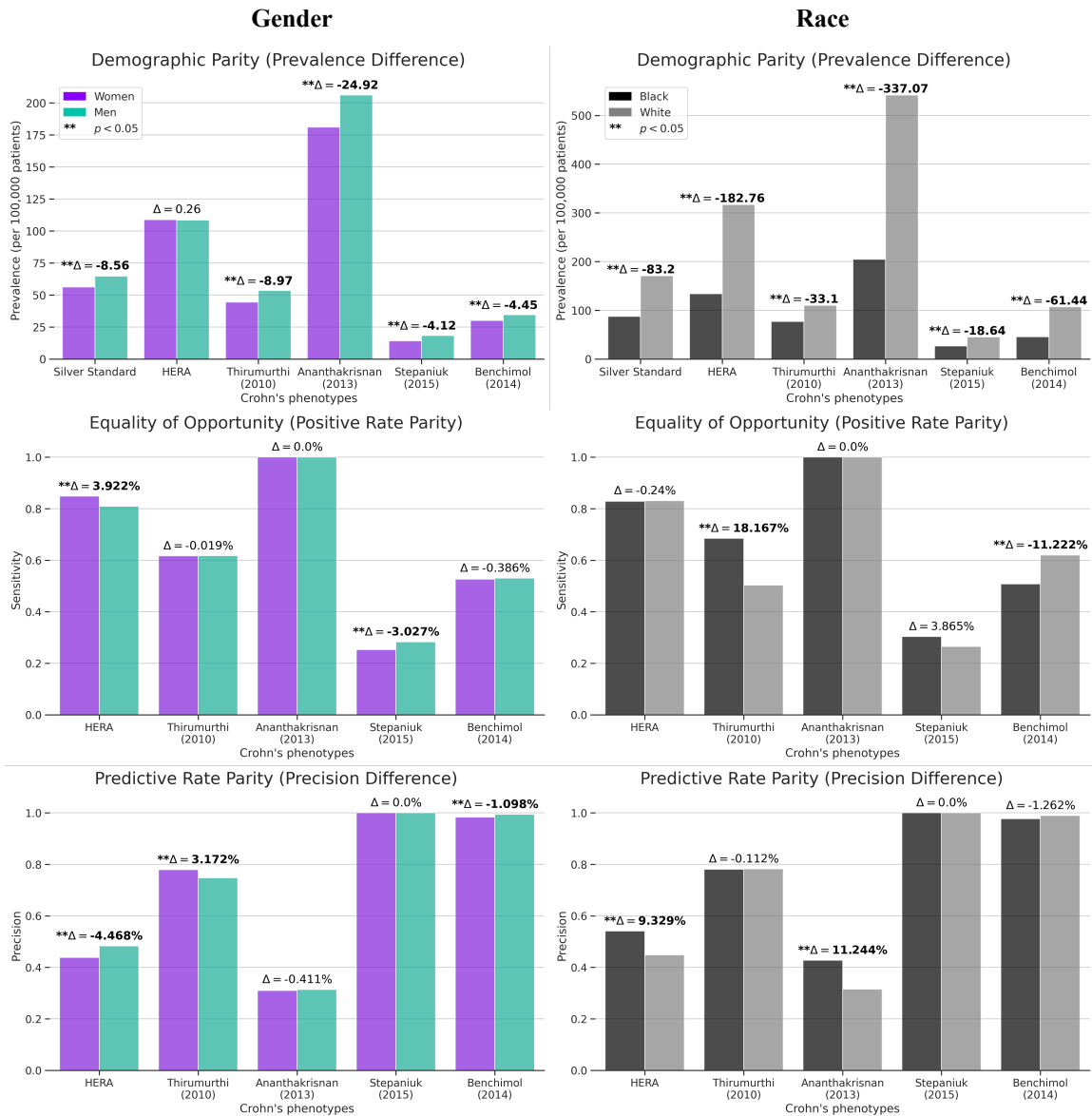

**Figure 3. Crohn's disease phenotype performance across gender and race for the five phenotypes measured across demographic parity, equality of opportunity, predictive rate parity.** A trade-off exists between using various phenotype definitions, as measured by the differences in performance across subgroups and fairness metrics

We visualize the demographic parity as the estimated prevalence difference between subgroups. The Crohn's disease phenotypes demonstrate a clear, consistent difference in estimated prevalence across subgroups, where the estimated prevalence is systematically higher among men than women, and among white patients than Black patients. This demographic disparity is statistically significant for all phenotypes for race, and all phenotypes besides HERA for gender. Recent Crohn's prevalence literature suggests that despite clinical manifestations of Crohn's disease being similar among Black and white patients[29], Black patients are often underdiagnosed and under-represented in Crohn's clinical trials[30], and that the true population prevalence for Black and white patients is approximately equal[31], suggesting that our data might also contain *access-to-care biases* favoring the inclusion of white patients.

The estimated sensitivity for the Crohn's phenotypes do not show a consistent pattern favoring any particular subgroup. However, the estimated sensitivity differences are noticeably impacted by hospitalization status. Phenotype definitions that require in-patient hospitalization or emergency room visits prior to Crohn's diagnosis, such as the Thirumurthi and Stepaniuk definitions, are particularly sensitive toward identifying men and Black patients, while other phenotypes, such as the HERA and Benchimol that included outpatient visits, are generally more sensitive for women and white patients respectively. This trend potentially reflects an *access-to-care bias*, where the healthcare setting impacts patient cohort inclusion. From these results, we contend men and Black patients at our hospital are more likely to be diagnosed or receive care for their Crohn's disease during an acute in-patient hospitalization, or during an emergency room visit, while women and white patients are more likely to be diagnosed for Crohn's during an out-patient visit. Existing literature highlights how Black children and adolescents with Crohn's disease are more likely to repeatedly visit the emergency room for disease management, highlighting how phenotype definitions requiring particular settings could be used to identify particularly at-risk patients[32].

The estimated precision for Crohn's phenotypes show an inconsistent pattern with significant differences across the various phenotypes. Among the race subgroups, the HERA and Ananthakrisnan definitions are more precise at identifying Black patients. When we consider gender subgroups, an interesting pattern emerges; unlike the other phenotype definitions, the HERA and Benchimol definitions require patients to have no prior history of Crohn's condition codes in their patient records before diagnosis and are more precise at identifying men. Meanwhile, the Thirumurthy definition that has no history requirement is more precise at identifying women. This disparate trend suggests that women are more likely to have received a previous (potentially relevant) diagnosis in their medical record compared to men, demonstrating a *diagnosis bias* where there are differences in initial presentation (e.g. women being more likely to have been suspected of having or previously having had the disease).

*Type 2 diabetes mellitus phenotypes*

The estimated prevalence for the diabetes phenotypes is consistently higher among Black patients than white patients. Across genders, two of the phenotype definitions (LEGEND, PheKB) estimate a higher prevalence for men, while one definition (Miller) estimates a higher prevalence for women. A study by Danaei et al. roughly estimates that type 2 diabetes mellitus is more prevalent among men than women[33]. Among Black and white patients, Signorello et al. suggest that after adjusting for socioeconomic status, prevalence rates across races are approximately equal[34]. Our results reflect the heterogeneous patient population at our academic medical center.

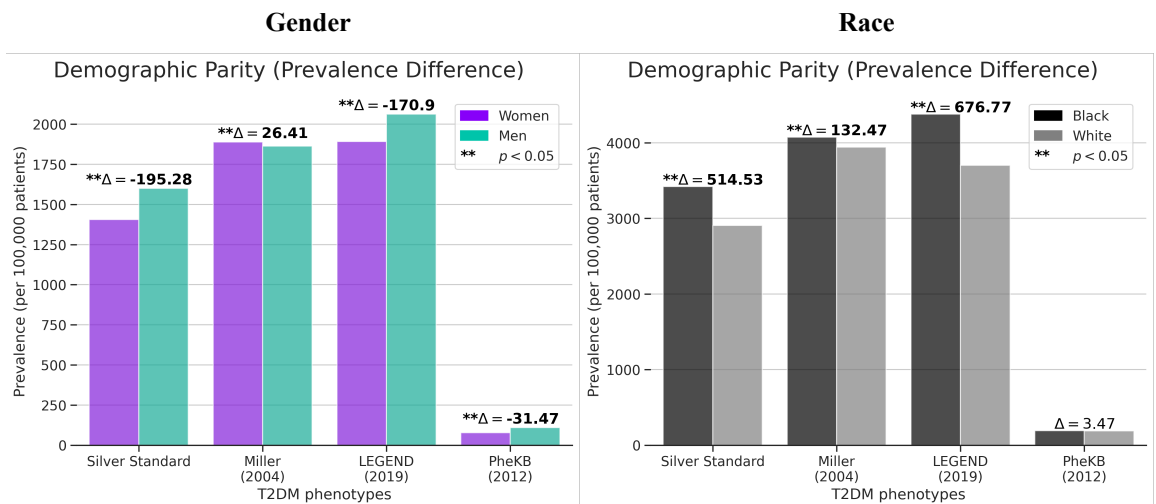

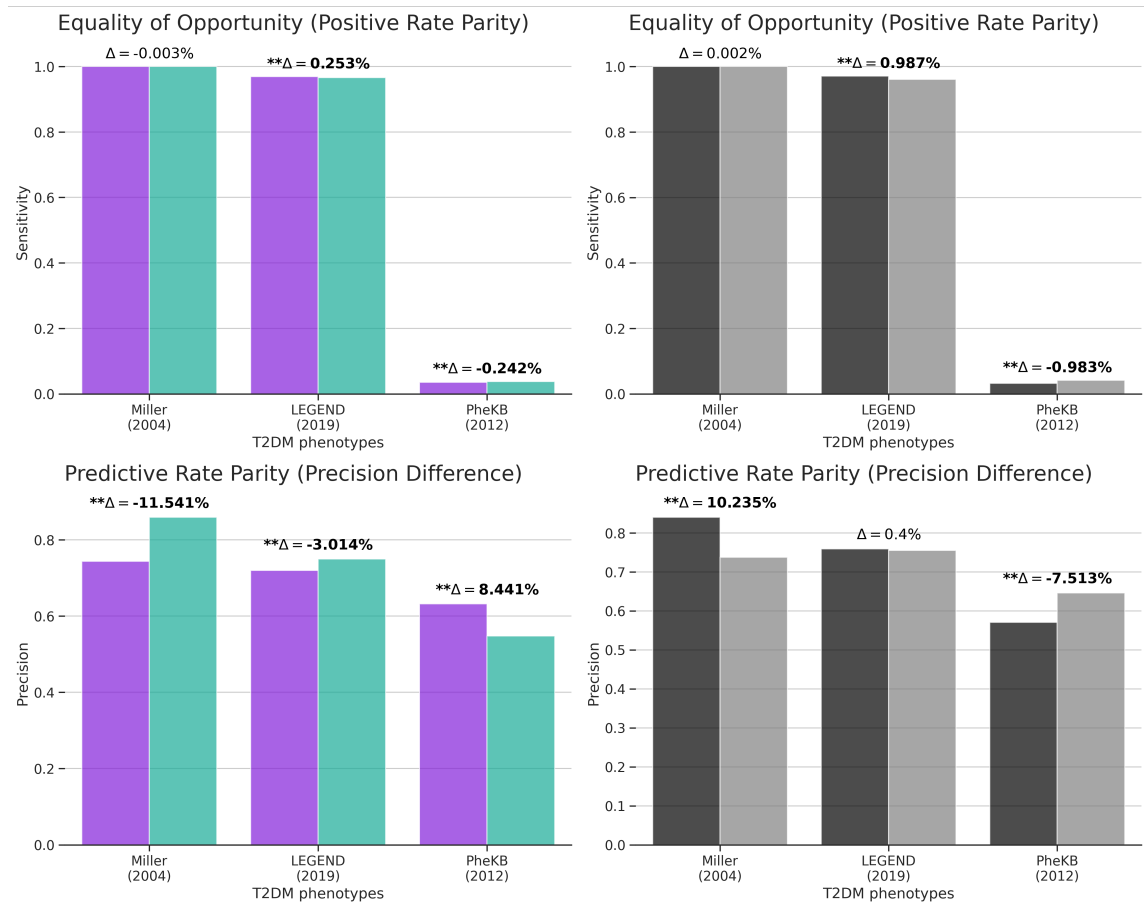

**Figure 4. Diabetes phenotype performance across gender and race for the three phenotypes measured across demographic parity, equality of opportunity, predictive rate parity.** A trade-off exists between using various phenotype definitions, as measured by the differences in performance across subgroups and fairness metrics

The estimated sensitivity for the diabetes phenotypes shows a mixed trend; LEGEND is more sensitive toward Black patients and women, while the PheKB definition is more sensitive toward white patients and men. When we critically examine the phenotype definitions (Table 3), the main differences are in medication requirements. All three definitions require patients to have been prescribed some form of medication, with the Miller definition including 18,368 drug concepts for "all glycemic meds" including insulin, while the LEGEND phenotype definition includes 18,290 drug concepts for "all type 2 diabetes mellitus medication excluding insulin". The PheKB definition is even more strict, including only 15,433 drug concepts for "type 2 diabetes prescriptions" that also excludes insulin. We note that the Miller definition (which includes the broadest category of medication, as well as insulin) performs approximately equally across genders and races. Thus, excluding insulin from the LEGEND and PheKB definitions potentially explains the difference in predicted sensitivities. This empirical analysis shows that future definitions of diabetes should potentially consider including insulin in its concept set if the research aims to include all patients diagnosed with type 2 diabetes mellitus.

The estimated precision for the diabetes phenotypes shows the Miller and LEGEND phenotypes having a higher precision for men and Black patients, while the PheKB phenotype having a higher precision for women and white patients. Given that the PheKB definition is the most stringent when it comes to inclusion and exclusion criteria, our results suggest that women and white patients are more likely to have received the more specific diabetes prescriptions that fit PheKB's narrower definition, while men and Black patients were more likely to have received insulin or other diabetes medication.

**Discussion and Conclusion**

In this study we present a workflow and best practices for assessing phenotypes using algorithmic fairness metrics, and apply our workflow to assessing Crohn's and diabetes phenotypes from literature. Our empirical analysis demonstrates that the appropriate use of a phenotype definition necessitates understanding the implicit biases generated during phenotype creation. We demonstrate this trade-off by measuring fairness metrics across protected categories, showing that even among highly-cited phenotype definitions that have influenced clinical guidelines, existing phenotype definitions can preferentially include (or exclude) certain protected subgroups. Because these trade-offs are unavoidable and researchers cannot avoid creating phenotype definitions, we advocate researchers (1) document, (2) open source, and (3) make transparent the process used to identify and protect or include particular subgroups of interest. By assessing phenotypes using fairness metrics these disparate impacts can be mitigated and balanced with the commensurate benefits to health. We highlight that the present study is only possible by the community's support of open science. It is only possible to conduct such assessments and improve the fairness of phenotypes because the authors have released their research as open source JSON definitions via GitHub. Finally, we encourage similar critique of our workflow and have released all analysis code. We encourage practitioners to extend our methods and report algorithmic fairness metrics in future observational health studies where health disparities might affect how a phenotype is defined or selected.